\documentclass[12pt,preprint]{iopart}

\usepackage{graphicx}
\usepackage{iopams}

\usepackage{xcolor}
\usepackage{isotope}

\begin{document}

\title{Self-Shielding Copper Substrate Neutron Supermirror Guides} 

\author{P. M. Bentley$^{1,5}$, R. Hall-Wilton$^{1,3,4}$,
  C. P. Cooper-Jensen$^1$, N. Cherkashyna$^1$, K. Kanaki$^1$,
  C. Schanzer$^2$, M. Schneider$^2$, P. B\"{o}ni$^2$}
\eads{phil.m.bentley@gmail.com, richard.hall-wilton@ess.eu,
  base548@kateandcarsten.com, Nataliia.Cherkashyna@ess.eu,
  kalliopi.kanaki@ess.eu, christian.schanzer@swissneutronics.ch,
  michael.schneider@swissneutronics.ch, peter.boeni@frm2.tum.de}

\address{
  $^1$ European Spallation Source ERIC, Box 176, 221 00 Lund,
  Sweden \\
  $^2$ Swiss Neutronics AG Neutron Optical Components \&
  Instruments, Br\"{u}hlstrasse 28, CH-5313 Klingnau, Switzerland \\
  $^3$ Universit\`{a} degli Studi di Milano-Bicocca, Piazza della Scienza 3, 20126 Milano, Italy \\
  $^4$ University of Glasgow, Glasgow, United Kingdom \\
  $^5$ Anderson Butovich Ltd. United Kingdom
}

\vspace{10pt}
\begin{indented}
\item[]\date{}
\end{indented}
\date{\today}

\begin{abstract}
  The invention of self-shielding copper substrate neutron guides is
  described, along with the rationale behind the development, and the
  realisation of commercial supply.  The relative advantages with
  respect to existing technologies are quantified.  These include ease
  of manufacture, long lifetime, increased thermal conductivity, and
  enhanced fast neutron attenuation in the keV-MeV energy range.
  Whilst the activation of copper is initially higher than for other
  material options, for the full energy spectrum, many of the isotopes
  are short-lived, so that for realistic maintenance access times the
  radiation dose to workers is expected to be lower than steel and in
  the lowest zoning category for radiation safety outside the
  spallation target monolith.  There is no impact on neutron
  reflectivity performance relative to established alternatives, and
  the manufacturing cost is similar to other polished metal
  substrates.
\end{abstract}

%
%
%
%
%

\section{Introduction}

With the imminent delivery of the first copper substrate neutron
guides to the European Spallation Source (ESS), currently under
construction in Sweden, it is timely to briefly report on the
rationale for, and development of, this new technology, and compare it
to established alternatives.

Neutron guides are solid tubes, usually of rectangular cross section.
The optimisation of the geometry of these devices is a very broad
topic and out of the scope of this paper. Here it is sufficient to
note that instrument angular resolution, transport efficiency and
background rejection often --- but not always --- tend to favour a
channel width in the region of 4-6\,cm, and due to mechanical
stability requirements the substrates themselves are typically around
1\,cm thick.

On the internal surfaces, neutron mirrors are deposited to maximise
the transmission of neutrons to the experimental stations.  This
allows the placement of the instrument remotely from the neutron
source, some 10s or even 100s of metres away, reducing instrumental
background and facilitating safe physical access designs.
Historically, single, smooth metallic layers were used as neutron
mirrors, but modern neutron guides use alternating Ni-Ti layers known
as ``supermirrors'' \cite{SUPERMIRRORS-MEZEI}.

In a typical neutron source, one has a means of neutron production
(fission, fusion, or spallation) and a volume of
temperature-controlled, strongly-scattering material known as a
moderator.  The optical interface between the neutron guides and the
region within a few metres of the moderator is often known as the
``beam extraction'' area, and specific technology exists to improve
the instrument performance.  The neutron optics work in the beam
extraction area carries greater technical challenges due to:
\begin{itemize}
\item The energy produced along with neutrons creating a large heat
  load
\item A high radiation environment
\item A need for good reflective properties at large angles near the
  source
\end{itemize}

High albedo, rad-hard materials are used around the moderator
(e.g. beryllium, with some recent interest in nanodiamond,
particularly for cold wavelengths \cite{NANODIAMOND-REFLECTOR}) where
the grazing angles are large.  At distances typically 1-2 metres from
the moderator, the grazing angles are below the critical angles on
supermirrors, but that still leaves heat- and radiation-loads that
must be carefully managed.

For the neutron guides entering the beam extraction area, polished
metal \cite{METAL-GUIDE-SUBSTRATES} and glass-ceramic
\cite{ZERODUR-SUBSTRATE} optics have existed for some time and offer a
robust, long-lifetime solution for high radiation environments.
Indeed, irradiation tests indicate that supermirrors on metallic
substrates do not show any degradation to a cumulative neutron fluence
of $9\times 10^{19}$\,$n$\,cm$^{-2}$ \cite{METAL-SUBSTRATE-LIFETIME}.
This is comparable to sodium float glass ($\sim 1\times
10^{20}$\,$n$\,cm$^{-2}$), and orders of magnitude higher than
boron-containing glass substrates ($\sim 1 \times
10^{18}$\,$n$\,cm$^{-2}$ for some borkron variants and $\sim 1 \times
10^{16}$\,$n$\,cm$^{-2}$ for borofloat).  These fluences are
integrated over the entire source spectrum, based on studies at the
reactor source of the Institut Laue-Langevin (ILL, Grenoble, France)
and the spallation source of the Paul Scherrer Institut (PSI,
Villigen, Switzerland).

They are assembled from shapes very similar to the glass guide
variants, but instead of bonding by adhesives they are bolted
together.  In addition to improved lifetime, they also have the
potential to enhance the fast neutron and gamma ray shielding
properties of the guide system, by increasing the density of the
material immediately outside the supermirror channel.  With the
preceding technology, there is always a gap of several centimetres
between the supermirror surface and the bulk shielding material, to
allow for glass substrates, adjustment, and vacuum housings.  The new
idea here is to bring dense shielding material into direct contact
with the supermirror.  This can correspondingly reduce the total
volume of shielding needed downstream through geometrical
considerations, by placing the most effective shielding in the key
locations where it will have the greatest impact.

It is important to note that the shielding effect is in the
longitudinal direction by virtue of the long line integral for low
divergence beams, and thus the advantages are in the far-field sense.
The transverse shielding effect is minimal, since the guide substrates
themselves are only $\sim$1\,cm thick and perhaps 2 metres of heavy
shielding is required in the beam extraction area; indeed locally
there should be an enhanced gamma production from neutron capture.

Further benefits of metal substrates over glass are improved thermal
conductivity, structural properties and robustness, which could allow
the guides to be placed close to the neutron source without thermal
damage occurring.  Of particular interest is using these for guide
inserts for in-monolith beam extraction from the target-moderator
region at spallation sources. However, regions further out where fast
neutrons are present may benefit from strategically placed effective
shielding as well.

Inspiration for the use of copper as a shielding material for fast
neutrons at spallation sources came in part from experience of its use
at other accelerator facilities and in high energy physics
experiments.  The possibility for its application here is motivated by
the same advantages: its shielding effect for fast neutrons, thermal
conductivity and structural properties.  A couple of examples are
given below:

In the Large Hadron Collider \cite{LHC} the area with the most intense
radiation environment is around the experimental interaction points
where the two beams collide.  To protect the delicate superconducting
accelerator equipment from such intense radiation and thermal loads,
there are two key protective elements: the TAS and the TAN.  The
Tertiary Absorber of Secondaries (TAS) is a 1.8m long block of copper
weighing around 2 tonnes and located at $\sim$20m from the interaction
point, which blocks high energy particles from exiting the
experimental cavern into the LHC tunnel; and the Tertiary Absorber of
Neutrals (TAN) \cite{LHCf}, which is a 3.5m long copper block,
designed to absorb neutral particles (neutrons and pions) at about
140m from the interaction point, which can have energies up to the
beam energy (7 TeV).

The hadronic calorimeter of the CMS instrument is made from brass
\cite{CMS-TDR}, whose stopping power for high energy particles
compares favourably with the steel calorimeter on ATLAS
\cite{ATLAS-HADRONIC-CALORIMETER}.  Unpublished concepts from JPARC
were also influential, where Cu had been used in the collimation of
instruments for the same reasons \cite{ARAI-PRIV}.

This manuscript looks at using copper for the multifunctional purpose
as substrate to the neutron supermirror guide, thermal transport and
radiation shielding.  Cost prevents deploying copper shielding
liberally, but in some targeted areas it would be ideal.

\section{Shielding at Spallation Sources}

The initial motivation for the copper guides, and copper shielding in
general, was to reduce fast neutron background signals on the
instruments of the ESS, based on investigations of challenges faced by
similar, operational facilities \cite{OVERCOMING-BACKGROUNDS}.
Shielding is a significant fraction of a neutron facilities cost, both
in terms of shielding the source and shielding the instruments.
Indeed, shielding and optics together represent half of the total cost
of constructing an instrument at a spallation source
\cite{BENTLEY-COST-OPTIMISATION}.

\begin{figure}
  \includegraphics[width=0.6\linewidth]{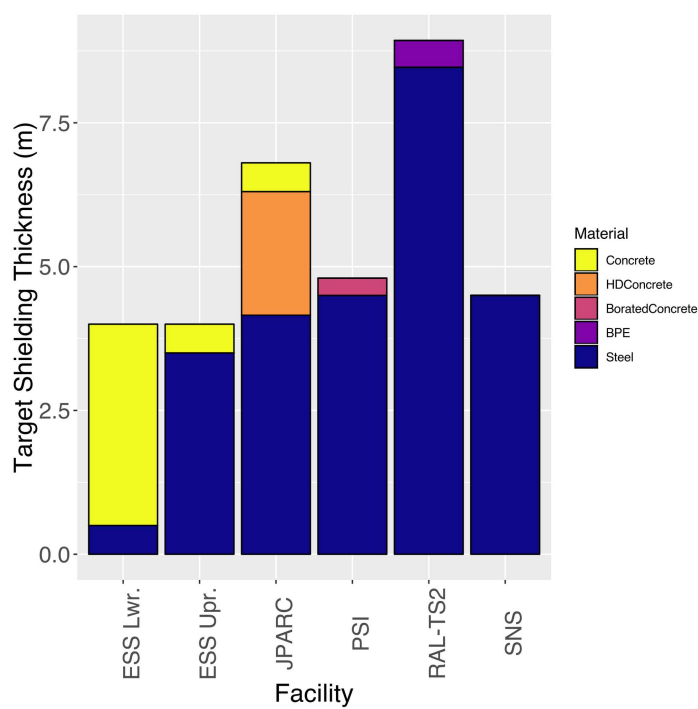}
  \caption{\label{fig:TargetShielding}Comparison of different
    spallation target shielding concepts at various leading facilities
    around the world, both operational and under construction.  The
    ESS has two concepts, a lower part and an upper part.
    A design change reduced the steel shielding for the lower parts
    physically below the neutron beam port level to save money.}
\end{figure}

\begin{figure}
  \includegraphics[width=0.6\linewidth]{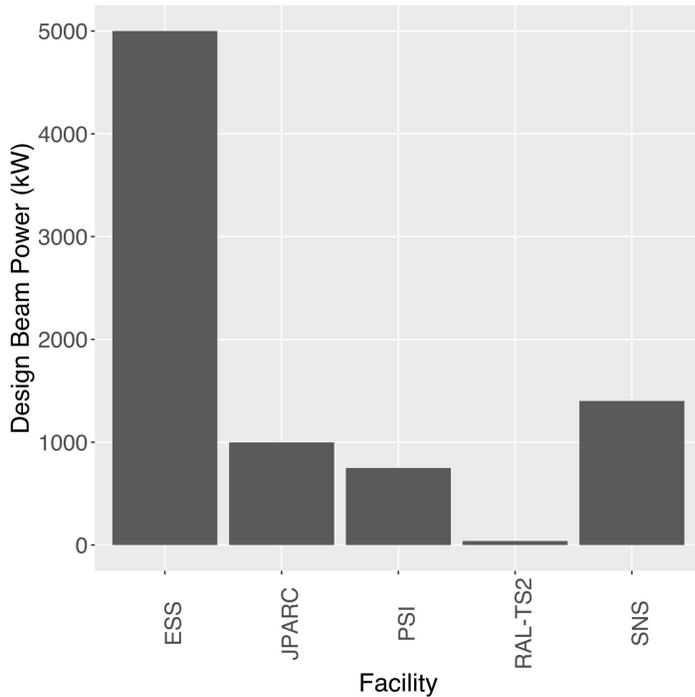}
  \caption{\label{fig:TargetPower}Comparison of different spallation
    target design powers at various leading facilities around the
    world, both operational and under construction.}
\end{figure}

Attempting to build the most powerful neutron spallation source in the
world, with the least amount of target shielding in the world, is a
challenging proposition from a background perspective.  Figures
\ref{fig:TargetShielding} and \ref{fig:TargetPower} give some insight
into the raw numbers of this challenge, and it should be no surprise
that Target Station 2 (TS2) at the Rutherford Appleton Laboratory
(RAL), in Oxford (UK) hosts the instruments with world-leading signal
to noise ratios (e.g. \cite{LET}).  This gives motivation to the quest for
better shielding.

The ESS Neutron Optics and Shielding group was formed to learn from
these other facilities and propose technical strategies related to
optics and shielding for the parts of the facility that are external
to the target system \cite{BENTLEY-COST-OPTIMISATION,NOSG-HANDBOOK}.
Collaboration with the ESS Detector group, who also had an interest in
reducing fast neutron backgrounds, brought the team to the realisation
that there were potential gains from exploring alternatives to steel
shielding in key locations \cite{NOSG-HANDBOOK}.  The need to minimise
fast neutron backgrounds is made especially poignant by the relatively
high sensitivity of thermal neutron detectors, especially
\isotope[3]{He}-based detectors, to fast
neutrons \cite{FN-eff1,FN-eff2,FN-eff3}.

\begin{figure}
  \includegraphics[width=0.4\linewidth]{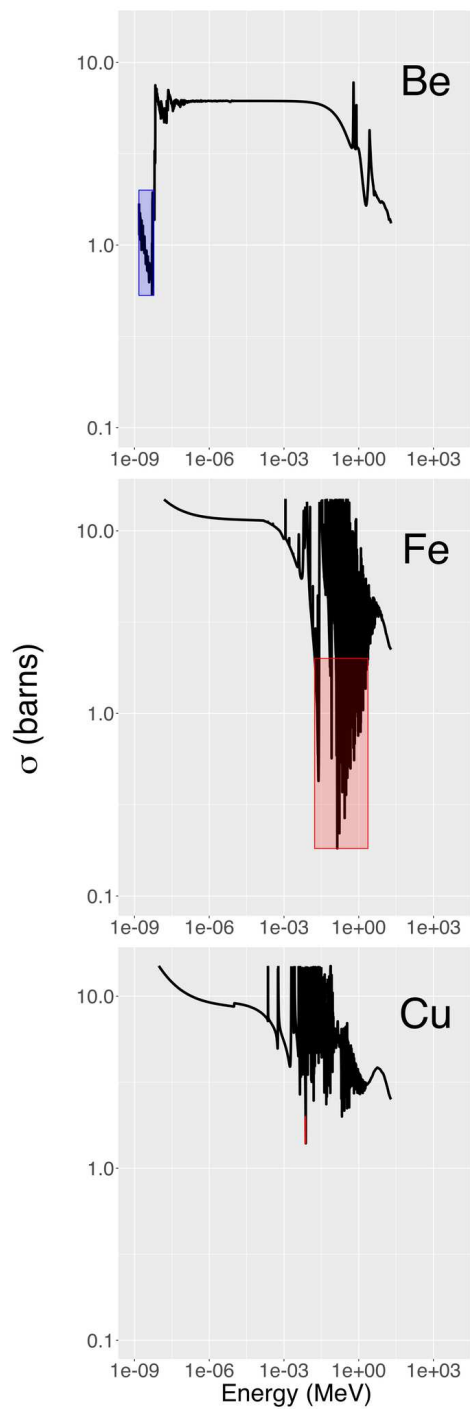}
  \caption{\label{fig:xsect} Total neutron cross section of Be, Fe and
    Cu \emph{vs} energy.  The Be data are spliced ENDF \cite{ENDF} and
    EXFOR \cite{EXFOR} data, whilst those for Fe and Cu are ENDF data
    only.  This is because the ENDF data for Be do not contain the
    transparent region at low energy.  The ``transparent'' region of
    the Be cross section is shaded with a blue box, with which one
    should compare the equivalent red box for Fe.}
\end{figure}

It is best practice to shield as close to the radiation source as one
can.  The primary area of concern is the beam-extraction area, however
one also needs to add heavy shielding near penetrations through bulk
shielding walls downstream, and regions where one aims to lose direct
line of sight to the source by curving the guides.  In these areas we
had the idea to improve the shielding properties of the optical
components, so that there was no gap between the supermirror and the
shielding material.  This means replacing glass and aluminium with
heavier elements.  In most shielding situations facing a high energy
neutron beam, under normal circumstances one would like to use steel
as the primary material.  Whilst it is possible to polish steel as a
substrate for supermirrors, the problem with the iron nucleus ---
known since the 1980s --- is the presence of resonance lines that act
as transparent windows, which may be linked to fast neutron background
problems \cite{CNCS-PROMPT-PULSE}.  To put this into perspective, the
cross section of Fe in parts of this energy range is comparable to
that of beryllium for cold neutrons.  Cooled Be filters are commonly
used neutron optical components \cite{BE-FILTER} to scatter out short
wavelength neutrons and transmit cold neutrons, thus one can visualise
steel as a filter that scatters thermal neutrons and, to some degree,
MeV--GeV neutrons, but transmits readily in the keV--MeV range with
only modest attenuation.  This is shown in figure \ref{fig:xsect},
highlighting the transparent regions.

There were a lot of myths in the community regarding these resonance
windows and fast neutron background effects, which had to be
systematically addressed at the outset.  For example, some proposed
using high-carbon steels as a general shielding material, with the
idea that the carbon atoms act as a moderator within the metal and
close the transparency windows.  ``High'' carbon steels are
$\sim$1\,\% carbon by weight, most steels are fractions of a \% carbon
by weight, so this has minimal effect.  Another myth was that the
background problem could be eliminated by increasing the counting time
of the experiment.  This confuses systematic error with statistical
error.  Fast neutron backgrounds vary as a function of time within the
data collection window, and furthermore they depend on the
configuration of the entire experimental suite and thus vary from
measurement to measurement \cite{SNS-DATA-MINING}, invalidating a
simple subtraction.  The \emph{raison-d'\^{e}tre} of high power
spallation sources is to improve measurements of weak phenomena that
are not possible at existing, lower power facilities, and such
measurements are disproportionately hindered by systematic and varying
background effects that resemble spurious signals.  Experience
elsewhere suggests that it is difficult to remove these effects and
they can impose significant operational limitations on measurements
with the instruments.

A final myth, and perhaps the most common, espouses the ignoring of
the problem entirely, relying on the idea that the fast neutron
background can be discriminated in time.  The underlying physics is
that of albedo reflections \cite{NEUTRON-ALBEDO} and photonuclear
processes ($\gamma \rightarrow n$ reactions) \cite{PHOTONUCLEAR} by
which fast neutrons are generated, and propagate down curved guides,
when no protons are illuminating the spallation target.  These two
effects combine to create long tailed distributions that exponentially
approach a flat asymptote between $10^{-2}$ and $10^{-4}$ relative to
the prompt $p+W \rightarrow n$ peak, as seen in operational
instruments at existing spallation sources
\cite{OVERCOMING-BACKGROUNDS,PMB-FAST-NEUTRON-DRAFT}.  The precise
asymptote level depends on the activation of the target, the geometry
of the Be, and to what degree the beamline shielding was designed to
impede fast neutron transport down the guide system.  The
characteristic width of the multi-exponential tail depends on the
spatial dimensions of the guide channels, the instrument cave, the
experimental hall, and the degree to which the shielding designs aim
to minimise fast neutron background signals.

In response to these identified risks, in 2013 the ESS teams
formulated a mitigation strategy \cite{NOSG-HANDBOOK}.  The definition
of this strategy, and the areas in which it was and was not followed,
are outside the scope of this article.  However, part of the core
strategy was to examine the use of superior shielding concepts close
to the neutron beamlines.

With steel as a primary shielding material, ideally one would like to
attach a second laminate layer, in the form of a hydrogenous material,
to deal with these resonant neutrons that shine through the primary
shielding.  Hydrogenous material is an effective moderator in the
energy range of the Fe resonance windows, and boron can be added to
conveniently absorb the thermalised neutrons that are emitted
downstream. The boron capture gamma is just under 0.5\,MeV which makes
it excellent in regards to subsequent photonuclear production.  For
this reason, borated polyethylene (BPE) and borated paraffin wax
feature heavily downstream of the primary steel shielding in the
neutronics design of TS2 at RAL.  At the Paul Scherrer Institut, in
Villigen (Switzerland), water tanks are used that are filled with
dilute boric acid.

Concrete is often used as a cheap alternative to borated hydrogenous
materials, with much compromised performance.  The water content and
density of concrete are rather low, so the cost savings are not
proportional.  For this reason, a separate project studied concrete
hydration \cite{CONCRETE-HYDRATION} and developed a modified recipe
for this purpose \cite{CARSTEN-CONCRETE,BPE-CONCRETE-GRAINS}.  This is
fine for instrument shielding and external shielding surfaces along
beamlines, but is not intended for shielding within a radius of a few
10s of cm from the neutron beamlines close to the source target to
reduce fast neutron contributions.

In the case of the ESS it was established that polymer solutions are
unsuitable for this use case, with a lifetime of only a few months
when the facility reaches its MW power classification
\cite{STUART-RUBBER-LIFETIME}.  This is because, as a long pulse
source, ESS may have a time-averaged neutron source brightness
comparable to the most powerful reactor source in the world --- the
Institut Laue-Langevin (ILL) in Grenoble, France --- significantly
exceeding anything offered by short pulse spallation sources.  For
this reason, copper is an attractive material that may offer
comparable shielding characteristics to a steel laminate solution,
without the short lifetime of polymers.  There is only one significant
resonance line creating a transparency in copper.  In early models of
simple bulk materials, copper provided more than an order of magnitude
suppression relative to iron for fast neutrons
\cite{CNCS-PROMPT-PULSE}.  Bags of copper shot in the gap between the
neutron guide and the shielding have subsequently been used to achieve
a 25\% reduction in the prompt pulse background
\cite{OVERCOMING-BACKGROUNDS} on the CNCS Instrument at SNS (ORNL),
Oak Ridge, US.

  \section{Method}

Being an item that is covered by nuclear licensing documentation, it
is important to specify the materials used in the preparation of the
substrate.  The coppers documented for use are CW008A and more
recently CW021A.  The substrate was cut, chamfered, polished and
cleaned according to the usual procedure at Swiss Neutronics AG.
Whilst copper is more expensive than aluminium as a raw material, the
preparation of copper is only slightly more challenging than other
materials.  Furthermore, the main cost drivers are subsequent to the
substrate preparation stages, so that there is a negligible economic
difference between the metal substrate options.  This simplifies the
derivation of a technical strategy for instrumentation
\cite{BENTLEY-COST-OPTIMISATION}, since only one metal substrate cost
needs to be considered, and the detailed design can specify the
substrate at a later stage.

The substrates were loaded onto the deposition facility in the normal
way, and an $m=3$ Ni-Ti supermirror was deposited according to the
standard commercial recipe used in other $m=3$ optical components.
Peel tests were passed, indicating good mirror adhesion equivalent to
other substrate materials.  Waviness measurements indicated no adverse
effects relative to other substrates, and the dimensional parameters
all lie within the usual engineering tolerances as are typical for
other materials.  

After fabrication, the supermirror samples were taken to the ILL and
the reflectivity curve measured on the D17 neutron reflectometer
\cite{D17-ILL}, and the Narziss neutron reflectometer at PSI, to
validate the quality of the mirrors.

Meanwhile, the shielding performance and activation of the materials
were assessed using Geant4 \cite{GEANT4,GEANT4b,GEANT4c} and
PHITS\,v3.17\,/\,DCHAIN \cite{PHITS} respectively, relative to other
materials.  The fast neutron attenuation between both these packages
was verified to be in agreement.  In a separately published study, as
part of the development programme, the shielding properties were
compared between Geant4 and neutron measurements at 174.1 MeV and
found to be in excellent agreement \cite{FAST-NEUTRON-MEAS}.

The activation calculations assumed a worker maintenance scenario.
Another unique design decision in the ESS was to remove heavy shutters
in the target monolith, as a cost saving measure.  Such beam shutters
are used regularly at existing neutron sources to turn off individual
beams for maintenance work without disturbing the operations of other
instruments outside of the work area.  The lack of such shutters means
that maintenance work at ESS will only be possible in the beam
extraction area during a complete shutdown of the facility, allowing
some additional time for removing the shielding.  It also means that
all of the technical staff will experience the same intensive burst of
activity every shutdown, with a strong motivation to access equipment
as early as possible in order to maximise the amount of repair during
the shutdown window.  Nonetheless, the unstacking of the shielding
bunker requires considerable effort and at least one day --- probably
longer.  

The calculations assumed an irradiation time of 5000 days, with the
ESS running at 5 MW proton beam power, using the least divergent,
0-1$^\circ$, cone of a source term generated from detailed spallation
target models that was later published as an internal reference
standard for the project \cite{ESS-SOURCE-TERM}.  This simulated
irradiation time is equivalent to 20 years of full power use assuming
annual cycles permitting 250 beam days.  Nine separate cooldown time
points were simulated after proton beam shutdown, these are: 0, 1, 6
hours and 1, 2, 4, 7, 14, 30 days.  The simulated starting material is
pure copper (CW008A). To give perspective, a range of other common
materials were compared that might be found close to the beam port at
spallation sources.

The location of the simulated material is outside the target monolith
at 6.5 m radius, to establish the activation conditions that
maintenance work may encounter with an unshielded guide component,
which is why the higher divergence columns can be ignored in the
source term.  A micro-kernel calculation was used to integrate the
total dose rate for a worker, assuming a 1 metre long guide section,
0.07 m tall, and a 0.4$\times$1\,m$^2$ human placed at 0.3\,m distance
from the material.  A more precise estimate may be required for a
specific, licensed task, however this estimate allows one to compare
materials adequately.

The PHITS/DCHAIN activation calculations were cross-checked against
independent MCNP5 \cite{MCNP} simulations of nuclear inventory
performed by the spallation target team \cite{LUCA-ZSOFIA-ACTIVATION},
and no noteworthy differences were found between the two sets of
results.

\section{Results and Discussions}

\subsection{Reflectometry}

\begin{figure}
  \includegraphics[width=0.6\linewidth]{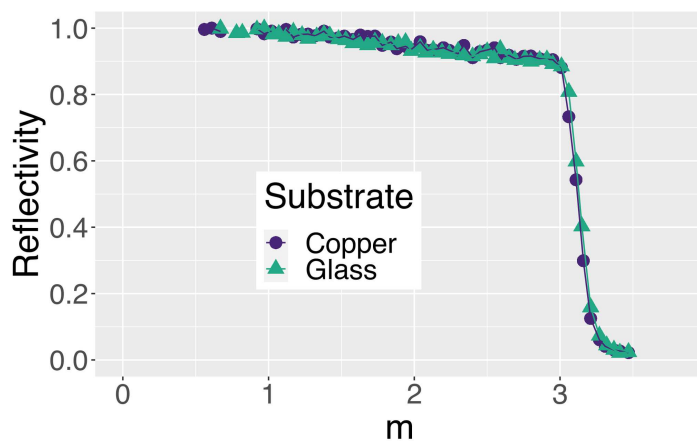}
  \caption{\label{fig:refl}Neutron reflectivity measurements of an
    $m=3$ supermirror on a production polished copper metal substrate,
    performed on the Narziss neutron reflectometer at PSI.
  }
\end{figure}

The reflectivity curve for the copper $m=3$ supermirror in production
is shown in figure \ref{fig:refl}, where $m=1$ corresponds to the
critical angle for total external neutron reflection of smooth nickel
($\approx$0.1$^\circ \times \lambda$, where the neutron wavelength
$\lambda$ is given in \AA{} units).  There one can see that the
neutron optical performance is excellent, and the copper substrate has
no effect at all on the supermirror --- as expected.

\subsection{Simulated Neutronic Properties}

\begin{figure}
  \includegraphics[width=0.6\linewidth]{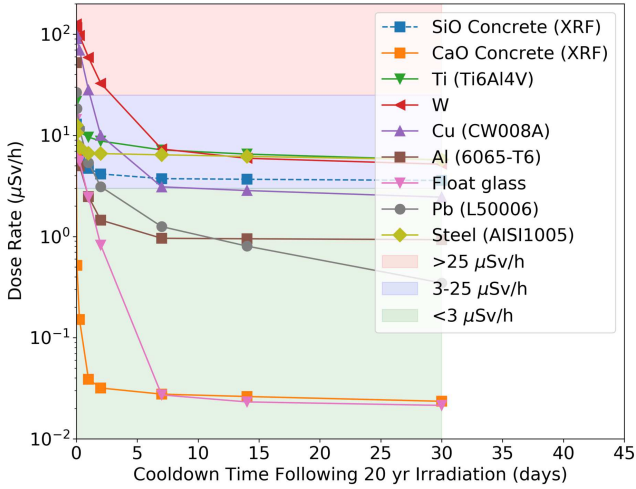}
  \caption{\label{fig:doseCurves} DCHAIN-based micro-kernel estimation
    of integrated dose rate of a human worker 0.3\,m from a
    1$\times$0.07\,$m^2$ guide element assuming copper, compared with
    other materials common to spallation source engineering.
  }
\end{figure}

The simulated dose rate to a worker, resulting from fast neutron
activation, is shown in figure \ref{fig:doseCurves}.  Here one sees
that for the first few days, copper is amongst the most active
materials, matching instrumentation experience with copper
monochromators, for example.  However, after $\sim$3\,days, the copper
dose rate drops below that of steel, and after approximately one week
copper would be expected to just enter the dose rate for ``green''
zone classification in Sweden ($<3\mu$Sv/h).  3--7 days of cooldown
time for maintenance access to guides close to a neutron source is not
unreasonable, based on experience at other facilities.  In the same
period, mild steel is around 6\,$\mu$Sv/h.  This is also not
problematic \emph{per se} but requires additional assessments, and one
should also bear in mind that the equipment inventory is quite dense
close to the spallation target structure, so one is not dealing only
with one activated source.  For this reason, best practice in nuclear
engineering is to use the ``ALARA'' principle, which stands for ``As
Low As Reasonably Achievable''.  Reducing the dose rate by 50\% for no
additional cost seems reasonable.

Figure \ref{fig:doseCurves} also shows other materials that are
interesting from an ESS shielding point of view.  Lead is an excellent
fast neutron shielding material, with an expected dose rate at all
times lower than copper and steel, and even below aluminium after
$\sim$12\,days.  As lead is also a good gamma attenuator, it makes
sense to line the front face of heavy shielding walls close to the
source with lead to both reduce the fast neutron activation downstream
and its resulting gamma shine, both improving the dose rate for
workers.

One can also see the benefit of replacing silicate aggregates with
limestone aggregates in concrete --- the input material composition of
both these concrete curves was generated from x-ray fluorescence (XRF)
measurements of concrete samples from a local
supplier.  This matches the
experience in Japan of Suzuki \emph{et al}
\cite{LIMESTONE-ACTIVATION}, for example.

The primary isotopes for medium-term management of fast-neutron
activated copper are summarised in table \ref{tab:isotopes}.
\begin{table}
  \caption{Table of most significant gamma contributors in copper,
    after 20 years of fast neutron activation 6.5 metres from the ESS
    target monolith followed by 30 days cooldown, determined using
    PHITS v3.17 and DCHAIN.}
  \label{tab:isotopes}
  \begin{indented}
    \item[]\begin{tabular}{l | r | r}
      \br
      Isotope & Dose Contribution (\%) & Half Life \\
    \mr
    $^{58}$Co & 31.5 & 2 mo\\
    $^{60}$Co & 27.0 & 5.3 yr\\
    $^{54}$Mn & 19.2 & 10 mo\\
    $^{48}$V  & 8.4  & 16 d\\
    $^{56}$Co & 6.0  & 77 d\\
    \br
    \end{tabular}
  \end{indented}
\end{table}
The three main isotopes contribute almost equally.  One expects
$^{58}$Co with a half life of 2 months, and $^{60}$Co with a half-life
of around 5.3 years, and $^{54}$Mn with a half life of 10 months.

These isotopes are generated in copper at a much lower rate than in
other shielding materials, even though they are similar to well known
nuclear waste problems.  For example, in steel, $^{54}$Mn is produced
by relatively common --- 70 millibarns (mb) ---
$^{54}$Fe(n,p)$^{54}$Mn fast neutron reactions and this accounts for
the overwhelming majority ($>$70\%) of the worker dose in the first
months after shutdown.  The copper channel is less than 1/4 of that,
at 16 mb, and comes from cascade processes.

Some stainless steels are known for rapidly accumulating dangerous
levels of $^{60}$Co, but this is from \emph{thermal} neutron
absorption in $^{59}$Co with a cross section of 37 barns, where cobalt
is typically a few \% by weight as a nickel impurity.  This process
vastly exceeds the copper production rate simulated here, with a
\emph{fast neutron} cross section of 14 mb, via the
$^{63}$Cu(n,$\alpha$)$^{60}$Co reaction.  Finally, $^{58}$Co
production occurs, again via high energy cascades, at 28 mb.  For an
overview of these higher energy isotope cross sections see the
following references
\cite{SISTERSON-FAST-NEUTRON-ACTIVATION,MEADOWS-FAST-NEUTRON-ACTIVATION,MICHEL-ACTIVATION,VANGINNEKEN-ACTIVATION}.

Reactor scientists who have worked with copper, as a monochromator on
a diffraction instrument for example, are used to the high levels of
activation in the first days.  Thermal neutron activation of copper
produces $^{64}$Cu, a beta emitter with a half life of just under 13
hours, decaying to $^{64}$Ni and $^{64}$Zn which are both stable.  The
second thermal activation isotope is $^{66}$Cu, decaying to $^{66}$Zn
with a half life of just over 5 minutes, which is stable.  These both
dominate in the early stages of cooldown, but the maintenance worker
dose from copper is significantly driven by only the spallation
products at higher energy, which have longer half lives.  It is for
these reasons that copper should be expected to allow arms' length
maintenance work after a few days of cooldown time, whilst steels fall
into the stricter radiation safety categories.  Neutron activation
should not be an impediment to the use of copper close to high power
spallation source beam lines.  It has less activity than steels, with
superior shielding performance of fast neutrons.

\subsection{Gamma Attenuation}

It is also worth briefly considering not just the direct fast neutron
attenuation but also the gamma attenuation improvements of copper
substrate guides, which can reduce downstream photonuclear
contributions and shielding load generally.  As shown in table
\ref{tab:gamma}, the half value layer (HVL) values indicate that
copper is slightly better than steel, but more importantly just over a
factor of 2 better than aluminium, with gains over glass expected to
be similar given the similar densities of glass and aluminium.

\begin{table}
  \caption{Gamma attenuation half value layer (HVL) of some example
    materials, for $^{60}$Co Gamma \cite{THORAEUS-GAMMA-ATTENUATION}.}
  \label{tab:gamma}
  \begin{indented}
  \item[]\begin{tabular}{l | r | r}
    \br
    Material & Density (g cm$^{-3}$) & HVL (mm) \\
    \mr
    \hline
    H$_2$O  &      1.0   &  108  \\
    Al    &      2.7   &  46.5 \\
    Fe    &      7.89  &  16.5 \\
    Cu    &      8.90  &  14.8 \\
    Pb    &      11.25 &  10.5 \\
    \br
  \end{tabular}
  \end{indented}
\end{table}

\subsection{Mechanical Factors}

The thermal conductivity of copper, compared to other metals, is found
in table \ref{tab:cond}.  There one sees that copper offers
substantial benefits for cooling possibilities in high power
spallation sources.
\begin{table}
  \caption{Thermal conductivity of some candidate substrate
    materials \cite{NIST-HEAT-DATABASE,THERMAL-CONDUCTIVITY-ELEMENTS}.}
  \label{tab:cond}
  \begin{indented}
  \item[]\begin{tabular}{l | r}
    \br
    Material & Thermal Conductivity (W m$^{-1}$ K$^{-1}$) \\
    \mr
    \hline
    Al    &      235  \\
    Fe    &       80  \\
    Cu    &      400  \\
    Glass &        $\sim$1  \\
    \br
  \end{tabular}
  \end{indented}
\end{table}
Whilst all of these materials are usable as substrates, the use of
copper facilitates the design process to meet the temperature
requirements of the monolith guide inserts by nature of its
significantly increased thermal conductivity.

\subsection{Production and Supply}

\begin{figure}
  \includegraphics[width=\linewidth]{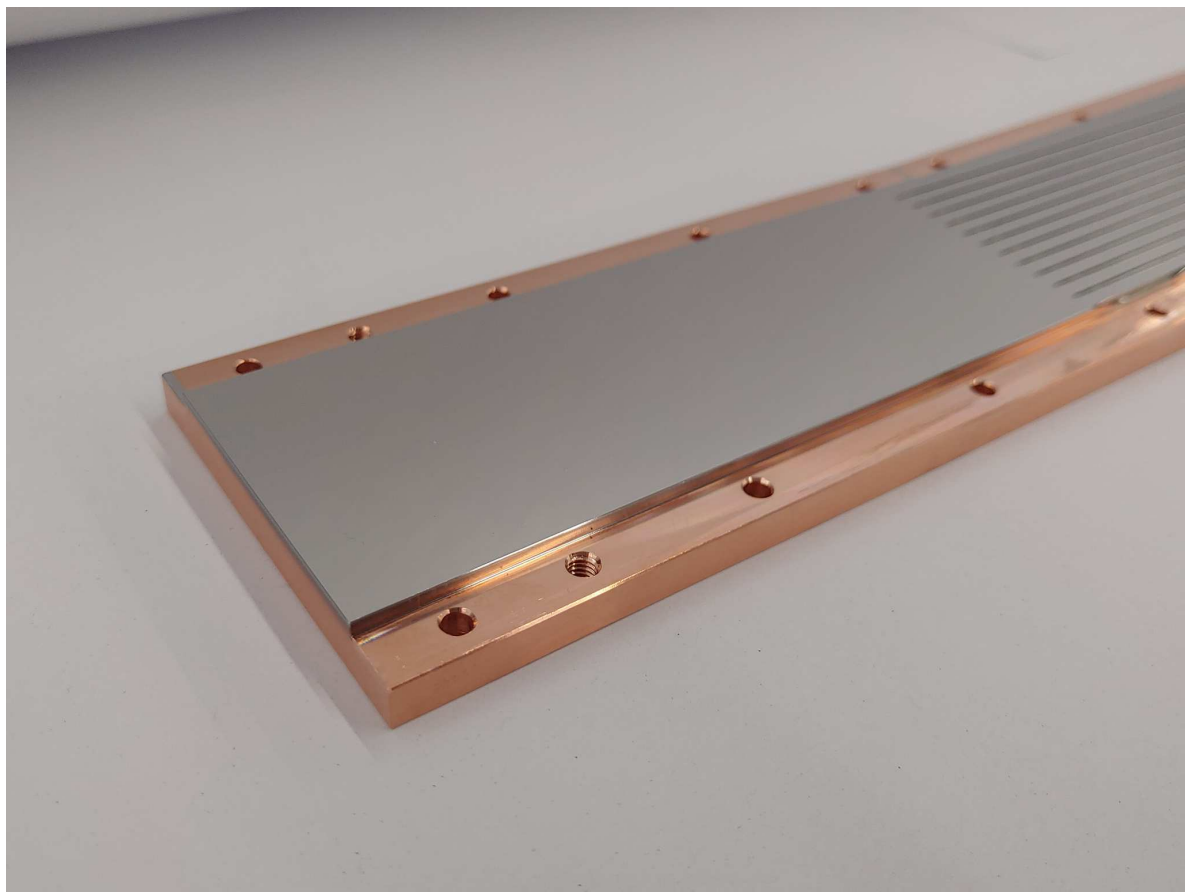}
  \caption{\label{fig:photo} Photograph of a supermirror deposited on
    a copper substrate, in preparation for assembly.  Also visible are
    bolt-holes for assembly and, to the right, grooves for inserting
    multi-channel optical wafers.}
\end{figure}

Copper substrates are now available in full industrial production.
The first units are arriving at ESS in Sweden for the VESPA instrument
at the time of writing, and the next unit is currently being
manufactured --- a photograph of one of those mirrors is shown in
figure \ref{fig:photo}, which also shows bolt-holes for assembly and
grooves for multi-channel optical wafers.  High $m$-value mirrors up
to $m=6$ have been produced.

All of the ESS instruments requiring supermirrors within the target
monolith are using copper substrates, at the time of writing there are
7 units at varying stages of procurement.  Several instruments are
also planning to order copper substrates in the near future, for some
of the key locations described earlier and in previous documentation
\cite{NOSG-HANDBOOK} along with copper shielding blocks.

\section{Conclusions}

Supermirrors with polished copper substrates have been developed and
are now available via routine commercial supply, at a similar cost to
other metal substrates.  They have been shown to offer excellent
optical performance, on par with any other substrate.  There are no
production issues expected to impact on longevity.

As part of an integrated concept of improved Cu-based shielding close
to high power spallation neutron beamlines, the combined developments
are expected to deliver significantly improved fast neutron
attenuation.  However, the ultimate background performance also
depends on holistic design across the facility, which is beyond the
scope of this article.

The main drawback is an enhanced activation level for the
first 3 days, compared to steel for example.  However, taking into
account \emph{realistic} maintenance access planning, this should not
be an issue.  Indeed, for long shutdowns --- the kind where
significant access would be expected in the beam extraction area
around spallation targets --- after a cooldown of $\sim$7 days, the
copper substrates themselves with 20 years of irradiation at the ESS
could allow maintenance work in the easiest safety category according
to Swedish legislation.  The actual access will most likely depend on
whether there are more active items nearby.  This is superior to all
steels under the same conditions.

Finally, copper offers a superior thermal conductivity compared to
other substrates, which facilitates the design of either active or
passive cooling systems with which to protect the supermirrors
themselves against thermal degradation.

\section{Acknowledgements}

S. Ansell, D. D. DiJulio and V. Santoro are gratefully acknowledged
for interesting scientific discussions that helped us with some of the
ideas in this work and other closely related papers.

\section{References}
\bibliographystyle{unsrt}
\bibliography{copperGuide}

\begin{thebibliography}{10}

\bibitem{SUPERMIRRORS-MEZEI}
F.~Mezei.
\newblock Novel polarized neutron devices: supermirror and spin component
  amplifier.
\newblock {\em Communications on Physics (London)}, 1(3):81--85, 1976.

\bibitem{NANODIAMOND-REFLECTOR}
V.~Nesvizhevsky, U.~K\"{o}ster, M.~Dubois, N.~Batisse, L.~Frezet, A.~Bosak,
  L.~Gines, and O.~Williams.
\newblock Fluorinated nanodiamonds as unique neutron reflector.
\newblock {\em Carbon}, 130:799--805, 2018.

\bibitem{METAL-GUIDE-SUBSTRATES}
C.~Schanzer, P.~B\"{o}ni, and M.~Schneider.
\newblock High performance supermirrors on metallic substrates.
\newblock {\em Journal of Physics: Conference Series}, 251:012082, 2010.

\bibitem{ZERODUR-SUBSTRATE}
C.~Lartigue, J.~R.~D. Copley, F.~Mezei, and T.~Springer.
\newblock Focusing of neutron beams using curved mirrors for small angle
  scattering.
\newblock {\em Journal of Neutron Research}, 5(1):71--79, 1996.

\bibitem{METAL-SUBSTRATE-LIFETIME}
C.~Schanzer, M.~Schneider, and P.~B\"{o}ni.
\newblock Metallic substrates for advanced applications in neutron optics.
\newblock In {\em Design and Engineering of Neutron Instruments (DENIM)
  Conference}, 2016.

\bibitem{LHC}
Lyndon Evans and Philip~Bryant (ed).
\newblock {LHC} machine.
\newblock {\em Journal of Instrumentation}, 3:S08001, 2008.

\bibitem{LHCf}
LHCf Collaboration.
\newblock The {LHCf} detector at the {CERN} large hadron collider.
\newblock {\em Journal of Instrumentation}, 3:S08006, 2008.

\bibitem{CMS-TDR}
G.~L. Bayatian et~al.
\newblock {CMS} the hadron calorimeter technical design report.
\newblock Technical report, {CERN}-{LHCC}-97-031 CMS-TDR-2, 1997.

\bibitem{ATLAS-HADRONIC-CALORIMETER}
A.~Succurro.
\newblock The {ATLAS} tile hadronic calorimeter performance in the {LHC}
  collision era.
\newblock {\em Physics Procedia}, 37:229--237, 2012.

\bibitem{ARAI-PRIV}
M.~Arai.
\newblock Private communication, 2015.

\bibitem{OVERCOMING-BACKGROUNDS}
N.~Cherkashyna, R.~J. Hall-Wilton, D.~D. DiJulio, A.~Khaplanov, D.~Pfeiffer,
  J.~Scherzinger, C.~P. Cooper-Jensen, K.~G. Fissum, S.~Ansell, E.~B. Iverson,
  G.~Ehlers, F.~X. Gallmeier, T.~Panzner, E.~Rantsiou, K.~Kanaki, U.~Filges,
  T.~Kittelmann, M.~Extegarai, V.~Santoro, O.~Kirstein, and P.~M. Bentley.
\newblock Overcoming high energy backgrounds at pulsed spallation sources.
\newblock {\em Proceedings of the 21st Meeting of the International
  Collaboration on Advanced Neutron Sources ({ICANS-XXI})}, 2014.

\bibitem{BENTLEY-COST-OPTIMISATION}
P.~M. Bentley.
\newblock Instrument suite cost optimisation in a science megaproject.
\newblock {\em J. Phys. Commun.}, 4:045014, 2020.

\bibitem{LET}
S.M.~Bennington R.I.~Bewley, J.W.~Taylor.
\newblock Let, a cold neutron multi-disk chopper spectrometer at isis.
\newblock {\em Nuclear Instruments and Methods in Physics A}, 637:128, 2011.

\bibitem{NOSG-HANDBOOK}
C.~Zendler, V.~Santoro, S.~Ansell, N.~Cherkashyna, D.~Martin Rodriguez,
  C.~Cooper-Jensen, D.~DiJulio, and P.~M. Bentley.
\newblock European spallation source neutron optics and shielding guidelines,
  requirements and standards.
\newblock Technical report, {ESS-0039408}, 2015.

\bibitem{FN-eff1}
F.~Piscitelli et~al.
\newblock Verification of {He}-3 proportional counters' fast neutron
  sensitivity through a comparison with {He}-4 detectors.
\newblock {\em Euro Physics Journal Plus}, 135:577, 2020.

\bibitem{FN-eff2}
G.~Mauri et~al.
\newblock Evidence of fast neutron sensitivity for $^3${He} detectors and
  comparison with boron-10 based neutron detectors.
\newblock {\em Euro Physics Journal Techniques and Instruments}, 6:3, 2019.

\bibitem{FN-eff3}
G.~Mauri et~al.
\newblock Fast neutron sensitivity of neutron detectors based on boron-10
  converter layers.
\newblock {\em Journal of Instrumentation}, 13:P03004, 2018.

\bibitem{CNCS-PROMPT-PULSE}
N.~Cherkashyna, K.~Kanaki, T.~Kittelmann, U.~Filges, P.~Deen, K.~Herwig,
  G.~Ehlers, G.~Greene, J.~Carpenter, R.~Connatser, R.~Hall-Wilton, and P.~M.
  Bentley.
\newblock High energy particle background at neutron spallation sources and
  possible solutions.
\newblock {\em Journal of Physics: Conference Series}, 528:012013, 2014.

\bibitem{BE-FILTER}
D.~C. Tennant.
\newblock Performance of a cooled sapphire and beryllium assembly for filtering
  of thermal neutrons.
\newblock {\em Review of Scientific Instruments}, 59:380, 1988.

\bibitem{ENDF}
M.~B. Chadwick et~al.
\newblock {ENDF/B-VII.1} nuclear data for science and technology: Cross
  sections, covariances, fission product yields and decay data.
\newblock {\em Nuclear Data Sheets}, 112:2887--2996, 2011.

\bibitem{EXFOR}
N.~Otuka et~al.
\newblock Towards a more complete and accurate experimental nuclear reaction
  data library ({EXFOR}): International collaboration between nuclear reaction
  data centres ({NRDC}).
\newblock {\em Nuclear Data Sheets}, 120:272--276, 2014.

\bibitem{SNS-DATA-MINING}
M.~B.~R. Smith, E.~B. Iverson, F.~X. Gallmeier, and B.~L. Winn.
\newblock Mining archived {HYSPEC} user data to analyze the prompt pulse at the
  {SNS}.
\newblock Technical report, {ORNL}/TM-2015/238, 2015.

\bibitem{NEUTRON-ALBEDO}
R.~C. Brockhoff and J.~K. Shultis.
\newblock A new approximation for the neutron albedo.
\newblock {\em Nuclear Science and Engineering}, 155:1--17, 2007.

\bibitem{PHOTONUCLEAR}
L.~Szilard and T.~A. Chalmers.
\newblock Detection of neutrons liberated from beryllium by gamma rays: a new
  technique for inducing radioactivity.
\newblock {\em Nature}, 134:494--495, 1934.

\bibitem{PMB-FAST-NEUTRON-DRAFT}
P.~M. Bentley.
\newblock Early operational challenges for the european spallation source in
  the sphere of shielding and optics.
\newblock {\em In preparation}, 2021.

\bibitem{CONCRETE-HYDRATION}
L.~Wads\"{o}, C.~P. Cooper-Jensen, and P.~M. Bentley.
\newblock Assessing hydration disturbances from concrete aggregates with
  radiation shielding properties by isothermal calorimetry.
\newblock {\em Physical Review Accelerators and Beams}, 20:043502, 2017.

\bibitem{CARSTEN-CONCRETE}
D.~D. DiJulio, C.~P. Cooper-Jensen, H.~Perrey, K.~Fissum, E.~Rofors,
  J.~Scherzinger, and P.~M. Bentley.
\newblock A polyethylene-{B}$_4${C} based concrete for enhanced neutron
  shielding at neutron research facilities.
\newblock {\em Nuclear Instruments and Methods in Physics Research A},
  859:41--46, 2017.

\bibitem{BPE-CONCRETE-GRAINS}
D.~D. DiJulio, C.~P. Cooper-Jensen, I.~Llamas-Jansa, S.~Kazi, and P.~M.
  Bentley.
\newblock Measurements and monte-carlo simulations of the particle
  self-shielding effect of {B}$_4${C} grains in neutron shielding concrete.
\newblock {\em Radiation Physics and Chemistry}, 147:40--44, 2018.

\bibitem{STUART-RUBBER-LIFETIME}
N.~Tsapatsaris and S.~Ansell.
\newblock Radiation dose and lifetime calculations for sealing materials used
  at neutron choppers at the {ESS}.
\newblock Technical report, European Spallation Source ESS-0084036, 2018.

\bibitem{D17-ILL}
R.~Cubitt and G.~Fragneto.
\newblock D17: the new reflectometer at the {ILL}.
\newblock {\em Applied Physics A}, 74:s329--s331, 2002.

\bibitem{GEANT4}
S.~Agostinelli et~al.
\newblock Geant4 - a simulation toolkit.
\newblock {\em Nuclear Instruments and Methods A}, 506:250--303, 2003.

\bibitem{GEANT4b}
J.~Allison et~al.
\newblock Geant4 developments and applications.
\newblock {\em {IEEE} Transactions on Nuclear Science}, 53:270, 2006.

\bibitem{GEANT4c}
J.~Allison et~al.
\newblock Recent developments in geant4.
\newblock {\em Nuclear Instruments and Methods A}, 835:186, 2016.

\bibitem{PHITS}
Tatsuhiko Sato, Yosuke Iwamoto, Shintaro Hashimoto, Tatsuhiko Ogawa, Takuya
  Furuta, Shin ichiro Abe, Takeshi Kai, Pi-En Tsai, Norihiro Matsuda, Hiroshi
  Iwase, Nobuhiro Shigyo, Lembit Sihver, and Koji Niita.
\newblock Features of particle and heavy ion transport code system ({PHITS})
  version 3.02.
\newblock {\em Journal of Nuclear Science and Technology}, 55(6):684--690,
  2018.

\bibitem{FAST-NEUTRON-MEAS}
D.~D. DiJulio, C.~P. Cooper-Jensen, H.~Bj\"{o}rgvinsd\'{o}ttir, Z.~Kokai, and
  P.~M. Bentley.
\newblock High-energy in-beam neutron measurements of metal-based shielding for
  accelerator-driven spallation neutron sources.
\newblock {\em Physical Review Accelerators and Beams}, 19:053501, 2016.

\bibitem{ESS-SOURCE-TERM}
V.~Santoro, D.~DiJulio, P.~M. Bentley, and L.~Zanini.
\newblock Source term for shielding design of bunker and beamlines at {ESS}.
\newblock Technical report, ESS-0416080, 2019.

\bibitem{MCNP}
MCNP.
\newblock https://mcnp.lanl.gov/.
\newblock Technical report, Los Alamos National Laboratory - LA-UR-03-1987,
  2003 (Revised 2008).

\bibitem{LUCA-ZSOFIA-ACTIVATION}
L.~Zanini.
\newblock Shielding of the neutron beam port inserts during extraction:
  baseline solution.
\newblock Technical report, European Spallation Source ESS-0094274, 2017.

\bibitem{LIMESTONE-ACTIVATION}
A.~Suzuki, T.~Iida, J.~Moriizumi, Y.~Sakamura, J.~Takada, K.~Yamasaki, and
  T.~Yoshimoto.
\newblock Trace elements with large activation cross section in concrete
  materials in {Japan}.
\newblock {\em Journal of Nuclear Science and Technology}, 38(7):542--550,
  2001.

\bibitem{SISTERSON-FAST-NEUTRON-ACTIVATION}
J.~M. Sisterson and J.~Ullmann.
\newblock Measurements of energy integrated cross sections for reactions
  producing relatively short-lived radionuclides using neutron beams with an
  energy range of 0.1–750 {MeV}.
\newblock {\em Nuclear Instruments and Methods in Physics Research B},
  234:419--430, 2005.

\bibitem{MEADOWS-FAST-NEUTRON-ACTIVATION}
J.~W. Meadows, D.~L. Smith, L.~R. Greenwood, R.~C. Haight, Y.~Ikeda, and
  C.~Konno.
\newblock Measured fast-neutron activation cross sections of {Ag}, {Cu}, {Eu},
  {Fe}, {Hf}, {Ni}, {Tb} and {Ti} at 10.3 and 14.8\,{MeV} and for the continuum
  neutron spectrum produced by 7\,{MeV} deuterons on a thick {Be}-metal target.
\newblock Technical report, Argonne National Laboratory ANL/CP-74797;
  CONF-9110261-3, 12 1991.

\bibitem{MICHEL-ACTIVATION}
R.~Michel, D.~Hansmann, S.~Neumann, W.~Glasser, H.~Schuhmacher, V.~Dagendorf,
  R.~Nolte, U.~Herpers, A.~N. Smirnov, I.~V. Ryzhov, A.~V. Prokofiev,
  P.~Malmborg, D.~Koll\'{a}r, and J.-P. Meulders.
\newblock Excitation functions for the production of radionuclides by
  neutron-induced reactions on {C}, {O}, {Mg}, {Al}, {Si}, {Fe}, {Co}, {Ni},
  {Cu}, {Ag}, {Te}, {Pb}, and {U} up to 180 {MeV}.
\newblock {\em Nuclear Instruments and Methods in Physics Research B},
  343:30--43, 2015.

\bibitem{VANGINNEKEN-ACTIVATION}
A.~Van Ginneken and A.~Turkevich.
\newblock Production of manganese 54 and zinc 65 from copper in thick targets
  by 0.45-{Gev}, 1.0-{Gev}, and 3.0-{Gev} protons.
\newblock {\em Journal of Geophysical Research}, 75:5121--5137, 1970.

\bibitem{THORAEUS-GAMMA-ATTENUATION}
R.~Thoraeus.
\newblock Attenuation of gamma radiation from $^{60}${Co}, $^{137}${Cs},
  $^{192}${Ir}, and $^{226}${Ra} in various materials used in radiotherapy.
\newblock {\em Acta Radiologica}, 3:81--86, 2009.

\bibitem{NIST-HEAT-DATABASE}
{NIST}.
\newblock {NIST} standard reference database 81 {NIST} heat transmission
  properties of insulating and building materials.
\newblock http://dx.doi.org/10.18434/T4RC7M, 1983.

\bibitem{THERMAL-CONDUCTIVITY-ELEMENTS}
C.~Y. Ho, R.~W. Powell, and P.~E. Liley.
\newblock Thermal conductivity of the elements.
\newblock {\em Journal of Physical and Chemical Reference Data}, 1(2):279,
  1972.

\end{thebibliography}

\end{document}